\documentclass{article}

  \usepackage[dandb, nonatbib, final]{neurips_2025}

\usepackage[utf8]{inputenc} 
\usepackage[T1]{fontenc}    
\usepackage{hyperref}       
\usepackage{url}            
\usepackage{booktabs}       
\usepackage{amsfonts}       
\usepackage{nicefrac}       
\usepackage{microtype}      
\usepackage{xcolor}         

\usepackage{makecell}
\usepackage{comment}
\usepackage{graphicx}

\title{Building Resilient Information Ecosystems: Large LLM-Generated Dataset of Persuasion Attacks}

\author{%
Hsien-Te Kao, Aleksey Panasyuk, Peter Bautista, William Dupree, \\{\bf Gabriel Ganberg, Jeffrey M. Beaubien, Laura Cassani, Svitlana Volkova}\\
  Aptima, Inc.\\
  Woburn, MA 01801 \\
}

\begin{document}

\maketitle

\begin{abstract}
Organization's communication is essential for public trust, but the rise of generative AI models has introduced significant challenges by generating persuasive content that can form competing narratives with official messages from government and commercial organizations at speed and scale. This has left agencies in a reactive position, often unaware of how these models construct their persuasive strategies, making it more difficult to sustain communication effectiveness. In this paper, we introduce a large LLM-generated persuasion attack dataset, which includes 134,136 attacks generated by GPT-4, Gemma 2, and Llama 3.1 on agency news. These attacks span 23 persuasive techniques from SemEval 2023 Task 3, directed toward 972 press releases from ten agencies. The generated attacks come in two mediums, press release statements and social media posts, covering both long-form and short-form communication strategies. We analyzed the moral resonance of these persuasion attacks  to understand their attack vectors. GPT-4’s attacks mainly focus on \textit{Care}, with \textit{Authority} and \textit{Loyalty} also playing a role. Gemma 2 emphasizes \textit{Care} and \textit{Authority}, while Llama 3.1 centers on \textit{Loyalty} and \textit{Care}. Analyzing LLM-generated persuasive attacks across models will enable proactive defense, allow to create the reputation armor for organizations, and propel the development of both effective and resilient communications in the information ecosystem.
\end{abstract}

\section{Introduction}
The news of public and private sector organizations plays a crucial role in ensuring transparency, fostering public trust, and communicating essential updates about policies, programs, and initiatives. It helps bridge the gap between complex oranizational functions and the public's understanding, making governance more accessible, accountable, and authentic \cite{martinelli2006strategic}. However, the rise of generative AI models, for example large language models (LLM) and multimodal frontier models like GPT4, CLaude, etc., has introduced significant challenges to the information ecosystem \cite{meier2024llm}. These models, with their ability to generate human-like text and images at scale, can be exploited to produce vast amounts of persuasive content, complicating the ability of organizations to maintain their messaging and ensure accurate representation of their public image~\cite{radivojevic2024human}. The speed and reach of such generative AI tools can amplify competing narratives, misinterpretations, or information overload undermining public confidence in official communications~\cite{chaudhary2024large}. This adversarial dynamic forces organizations to not only enhance the clarity and consistency of their messages, but also adapt to an evolving information environment where their narratives are constantly at risk of being distorted or overshadowed.

The diverse range of persuasive techniques that LLMs can employ makes them highly effective in shaping perceptions and influencing discourse on a broad scale. These models can frame information in ways that offer overly simplistic explanations, sometimes overlooking the nuances of a situation, which can lead to misunderstandings or oversimplified interpretations of complex issues \cite{breum2024persuasive}. At times, the discourse shifts focus away from critical discussions or vital details, diverting attention to less relevant points, thereby weakening the original argument \cite{dontcheva2020persuasive}. In addition, persuasive tactics may lean on emotional appeals or loaded language, steering the reader's opinion through careful word choice, rather than offering a balanced, reasoned perspective \cite{bassi2024decoding}. In certain instances, the credibility of individuals or entities can be questioned through subtle or overt attacks, undermining their authority, and dismissing valid concerns \cite{hughes2014discrediting}. This challenge requires agencies to refine their strategies and navigate an increasingly complex information landscape, ensuring that their messages maintain clarity and resonance amid competing narratives~\cite{NATOInfoThreat2024, NATODynamicResilience2025}.

The persuasive power of LLMs is rooted in the targeting of core human values. Using fundamental beliefs, LLMs can frame arguments in ways that resonate with deep-rooted instincts, such as a sense of harm or injustice, often eliciting changes in perspective from the audience \cite{carrasco2024large}. At the same time, generative models leverage societal notions of hierarchy and respect for authority, influencing how information is presented to align with established norms \cite{hutchens2023language}. LLMs can draw on collective feelings of solidarity, reinforcing group identity and encouraging behaviors that prioritize the collective over the individual \cite{mylrea2025generative}. In addition, they can appeal to ideals of purity and discipline, suggesting that certain actions or beliefs align with higher moral standards, which can be particularly effective in shaping public opinion on complex issues \cite{matz2024potential}. However, organizations continue to remain in the dark about how these models are crafting competing narratives that subtly exploit deeply ingrained moral principles across persuasion attacks e.g., Appeal to Emotions or Authority, making it difficult for them to anticipate and respond to their influence.

The persuasive nature of LLM-generated content has left organizations largely reactive, with little insight into how LLMs leverage  persuasive tactics. In this paper, we introduce a large LLM-generated persuasion attack dataset, which contains open and closed LLMs -- GPT-4, Gemma 2, and Llama 3.1 generated persuasive attacks on government agency press releases. The dataset includes 134,136 attacks across 23 persuasive techniques from SemEval 2023 Task 3, targeting 972 news releases from ten agencies. We also examined the moral resonance of these attacks to better understand their persuasive appeals. GPT-4's attacks mainly resonate with \textit{Care}, emphasizing protection, while also incorporating \textit{Authority} and \textit{Loyalty}, reinforcing trust in leadership and group cohesion. Gemma 2's attacks focus on \textit{Care}, prioritizing concern and protection, and also leverage \textit{Authority} for trust in leadership. Llama 3.1's attacks primarily emphasize \textit{Loyalty}, highlighting group identity and tradition, with \textit{Care} also playing a strong role in evoking emotional responses. While all models emphasize \textit{Care}, GPT-4 and Gemma 2 rely more on \textit{Authority}, and Llama 3.1 focuses more on \textit{Loyalty} to shape their persuasive appeals.

\section{Related Work}
Recent research highlights the significant role of LLMs in shaping public discourse. LLMs generate persuasive content at scale, framing public conversations and steering information in particular directions \cite{garry2024large,NATOInfoThreat2024}. LLMs can be easily prompted to create and complete information narratives, allowing users to generate localized content that aligns with specific objectives, reaching targeted audiences with minimal effort \cite{williams2024large}. The ability of LLMs to generate content indistinguishable from human writing complicates the identification of narrative sources, blurring the boundaries between generated and human content \cite{guo2024online}. As LLMs generate messaging campaigns at scale, collective intelligence can be compromised, as these models dominate discussions and shift focus toward content created with specific goals in mind, overshadowing diverse and genuine human input \cite{burton2024large}. The structured process of messaging generation, from creation to dissemination, shows how LLMs can shape strategic discourse, amplifying certain perspectives while minimizing others \cite{grovs2024information,NATOCognitiveWarfare2023}. As LLMs evolve, addressing their influence on discourse requires new proactive strategies to mitigate risks of generated content shaping the informartion environment.

Persuasive techniques significantly impact text-based communications across media. SemEval-2021 Task 6 expanded the persuasion taxonomy from 20 to 22 techniques, emphasizing additional visual-specific techniques in both texts and images for multimodal persuasion \cite{dimitrov2021semeval}. SemEval-2023 Task 3 broadened the taxonomy from 20 to 23 techniques, highlighting the importance of text-based framing and genre classification in news articles \cite{piskorski2023semeval}. SemEval-2024 Task 4 introduced a hierarchical categorization of 22 techniques, providing a clearer framework for more detailed and accurate evaluation in memes \cite{dimitrov2024semeval}. Research into online posts shows that emotional appeals and repetition play key roles in shifting discourse and completing narratives within online contents \cite{chen2021persuasion}. Additionally, digital environments influence public discourse by framing perceptions through collective social practices, particularly in online spaces \cite{foster2023truth}. These studies highlight the variety of persuasive techniques, emphasizing the need for comprehensive understanding of their diverse nature and how they manifest in different contexts.

Moral foundations are central to persuasion and discourse across various topics, shaping how arguments are formed and received. Research shows that moral reframing—adapting arguments to align with the audience’s moral values—enhances persuasion, especially when values differ \cite{feinberg2019moral}. Moral appeals are more effective when aimed at individuals whose beliefs are rooted in moral considerations, highlighting the importance of understanding the audience's moral framework \cite{kodapanakkal2022moral}. These arguments not only persuade but also strengthen moral convictions, making individuals less likely to compromise or entertain opposing views \cite{luttrell2019challenging}. Additionally, perceiving a moral basis for attitudes increases consistency between beliefs and actions, boosting resistance to persuasion from alternative perspectives \cite{harre1985persuasion}. Persuasion is also shaped by authoritative figures, such as scientists and research organizations, who use rhetorical strategies to influence public discourse \cite{alshomary2022moral}. Together, these findings highlight the power of moral framing and the need to understand persuasive techniques through the lens of moral foundations to navigate communication effectively.

Existing research has highlighted the influence of LLM-generated content on public discourse, particularly regarding their persuasive capabilities. However, there remains a critical gap in understanding the specific persuasive techniques used by LLMs, particularly in government communications. While persuasive strategies of LLMs are explored at a broad level, there is a lack of detailed, model-specific analysis of how different LLMs like GPT-4, Gemma 2, and Llama 3.1 deploy distinct persuasive tactics. Furthermore, much of the existing work does not systematically examine the variety of persuasive techniques these models employ, especially in contexts that significantly impact public policy and communication. Our work bridges these gaps by offering a large LLM-generated persuasion attack dataset, which consists of 134,136 comprehensive persuasive attacks generated by leading LLM models, covering 23 persuasive techniques and 972 news articles from 10 agencies. By examining the underlying moral resonance, we offer insights that enrich the understanding of LLM-generated persuasive attacks that support the development of proactive communication.

\section{Agency Press Releases}
We collected 972 press release articles from ten agencies, each with 100 articles except for IARPA, which had only 72 at the time of collection. These include Air Force Research Laboratory (AFRL), Army Research Laboratory (ARL), Defense Advanced Research Projects Agency (DARPA), Defense Innovation Unit (DIU), Department of Defense (DoD), Defense Threat Reduction Agency (DTRA), Intelligence Advanced Research Projects Activity (IARPA), National Geospatial-Intelligence Agency (NGA), Naval Research Laboratory (NRL), and Sandia National Laboratories (SNL). 

\paragraph{Domain Labeling}
We labeled each article into multiple domains and topics using GPT-4o Mini 2024-07-18 \cite{achiam2023gpt}. The top three domains for each agency is as follows: AFRL articles predominantly fell under Defense (90\%), R\&D (51\%), and Technology (50\%). ARL articles were categorized as Defense (88\%), Technology (62\%), and R\&D (57\%). DARPA's coverage was primarily in Defense (97\%), Technology (80\%), and R\&D (39\%). DIU articles were overwhelmingly focused on Defense (98\%) and Technology (93\%), with Aerospace (29\%). DOD articles were classified mainly under Defense (99\%), International Relations (65\%), and Government (29\%). DTRA’s content was centered on Defense (97\%), International Relations (58\%), and Security (25\%). IARPA’s coverage included Technology (79\%), Defense (42\%), and Intelligence (40\%). NGA articles were associated with Defense (61\%), Geospatial Intelligence (55\%), and Government (39\%). NRL content was categorized under Defense (86\%), R\&D (60\%), and Technology (48\%). Lastly, SNL’s articles primarily focused on R\&D (68\%), Technology (64\%), and Engineering (40\%).

\paragraph{Topic Labeling}
The top three topics for each agency are as follows: AFRL articles were categorized under Military Technology (12\%), Military Operations (8\%), and R\&D (8\%). ARL’s coverage included Military Technology (21\%), Machine Learning (14\%), and AI (11\%). DARPA articles were classified under Machine Learning (12\%), National Security (10\%), and Military Technology (8\%). DIU’s content was focused on Defense Innovation (21\%), AI (14\%), and Military Technology (13\%). DOD articles were associated with Military Cooperation (11\%), Regional Security (9\%), and Crisis Management (9\%). DTRA’s coverage centered on WMD (12\%), Threat Reduction (12\%), and Counterterrorism (7\%). IARPA articles were categorized under AI (22\%), Data Analysis (15\%), and Machine Learning (11\%). NGA’s content primarily fell under Geospatial Intelligence (22\%), National Security (14\%), and Geospatial Analysis (14\%). NRL articles were classified under Materials Science (9\%), Remote Sensing (8\%), and Naval Research (8\%). Lastly, SNL’s coverage included STEM Education (13\%), Materials Science (13\%), and Nanotechnology (12\%).

\section{Persuasion Attack Generation}
Due to ethical considerations, we will not disclose the exact prompt used for attack generation, as doing so could lead to negative implications, potentially enabling harmful applications that undermine the intent of this paper. Instead, we provide a high-level structural overview that facilitates responsible technical discussion. The structure of the prompt is intentionally layered, with each element contributing to a cohesive and impactful critique. The "Attack Fallacy – Description" outlines the persuasive technique and its details, while the "Persona" and "Perspective Guideline" components define the LLM's mindset toward the new articles, focusing on the implications. The "Argument Guideline" introduces a logical structure that strictly aligns with the single attack provided, and the "Writing Style Guideline" establishes a strong, forceful tone while eliminating distractions and irrelevant elements. Together, these components form a well-defined blueprint for generating skeptical, logical, and impactful persuasive attacks on agency press releases, ensuring the critique remains disciplined, purposeful, and methodologically sound.

We utilized three LLMs (GPT-4 0613 \cite{achiam2023gpt}, Gemma 2 9B Instruct \cite{team2024gemma}, and Llama 3.1 8B Instruct \cite{grattafiori2024llama}) to generate 23 persuasive attacks from SemEval 2023 Task 3 for each of the 972 news articles from ten agencies. Each persuasive attack is crafted in two distinct mediums (press releases and social media posts) to capture both long-form and short-form communication. This dual-medium approach ensures a comprehensive representation of persuasive attacks in different communication channels. All generated persuasive attacks are framed from a harmful perspective, targeting the key points of the news articles to diminish the overall messaging. In total, we generated 134,136 persuasive attacks, both locally hosted and through API calls, for the \href{https://github.com/Aptima/llm-generated-persuasion-attack-dataset/tree/main}{large LLM-generated persuasion attack dataset}. The following are the 23 persuasive attacks used in the attack generation: \textit{Appeal to Authority}, \textit{Appeal to Popularity}, \textit{Appeal to Values}, \textit{Appeal to Fear}, \textit{Flag Waving}, \textit{Causal Oversimplification}, \textit{False Dilemma}, \textit{Consequential Oversimplification}, \textit{Straw Man}, \textit{Red Herring}, \textit{Whataboutism}, \textit{Slogans}, \textit{Appeal to Time}, \textit{Conversation Killer}, \textit{Loaded Language}, \textit{Repetition}, \textit{Exaggeration}, \textit{Obfuscation}, \textit{Name Calling}, \textit{Doubt}, \textit{Guilt by Association}, \textit{Appeal to Hypocrisy}, and \textit{Questioning the Reputation}.

\section{Moral Resonance Analysis}
Moral Foundations Theory (MFT) is a psychological framework explaining the moral values shaping human judgment and behavior \cite{graham2013moral}. These moral foundations play a crucial role in persuasion, as effective communication aligns with an audience’s deeply held values to enhance receptivity and influence decision-making \cite{hurst2020messaging}. MFT identifies five core foundations: \textit{Care}, capturing compassion, empathy, and the desire to prevent harm; \textit{Fairness}, encompassing justice, equality, and proportionality in exchanges; \textit{Loyalty}, reflecting group belonging, commitment, and allegiance to one’s community; \textit{Authority}, relating to respect for tradition, leadership, and social order; and \textit{Purity}, pertaining to sanctity, self-discipline, and avoidance of contamination, physical and moral. These foundations contribute to moral resonance, where individuals feel deep emotional or cognitive alignment with messages affirming their identity and beliefs \cite{giorgi2017mind}. Persuasive communication leverages moral resonance by framing narratives, arguments, and policies in ways strongly reflecting an audience’s foundational moral intuitions, reinforcing beliefs and motivating action. For each LLM-generated persuasive attack, we used the moral foundations dictionary \cite{ji2020developing} to identify virtue words for each moral foundation and calculate the ratio of words associated with each foundation.

\subsection{GPT-4 Findings}
\begin{table}
\footnotesize
  \caption{\textbf{Moral resonance of GPT-4 generated attacks.} These attacks primarily align with \textit{Care}, followed by \textit{Authority} and \textit{Loyalty}, emphasizing emotional appeal, trust in hierarchy, and group cohesion. In contrast, \textit{Fairness} and \textit{Purity} show significantly weaker resonance, indicating that its persuasive strategies are less focused on justice and moral sanctity.}
  \label{gpt4-table}
  \centering
  \resizebox{0.87\textwidth}{!}{
  \begin{tabular}{lcccccc}
    \toprule
    \textbf{Name} & \textbf{Care} & \textbf{Fairness} & \textbf{Loyalty} & \textbf{Authority} & \textbf{Purity} & \textbf{Resonance} \\
    \midrule
    Appeal to Authority & 0.21 & 0.11 & 0.22 & \textbf{0.42} & 0.03 & Authority \\
    Appeal to Popularity & 0.37 & 0.09 & 0.25 & 0.24 & 0.04 & Care \\
    Appeal to Values & 0.29 & 0.10 & 0.26 & 0.29 & 0.06 & Duo \\
    Appeal to Fear & 0.31 & 0.12 & 0.27 & 0.26 & 0.04 & Care \\
    Flag Waving & \textbf{0.38} & 0.13 & 0.23 & 0.22 & 0.03 & Care \\
    \makecell[l]{Causal \\ Oversimplification} & 0.31 & \textbf{0.28} & 0.15 & 0.22 & 0.03 & Care \\
    False Dilemma & 0.28 & 0.14 & 0.29 & 0.24 & 0.05 & Loyalty \\
    \makecell[l]{Consequential \\ Oversimplification} & 0.35 & 0.11 & 0.24 & 0.25 & 0.05 & Care \\
    Straw Man & 0.33 & 0.12 & 0.24 & 0.27 & 0.04 & Care \\
    Red Herring & 0.31 & 0.11 & 0.28 & 0.26 & 0.04 & Care \\
    Whataboutism & 0.29 & 0.10 & 0.30 & 0.27 & 0.05 & Loyalty \\
    Slogans & 0.30 & 0.13 & 0.24 & 0.30 & 0.03 & Duo \\
    Appeal to Time & 0.30 & 0.09 & \textbf{0.40} & 0.19 & 0.02 & Loyalty \\
    Conversation Killer & 0.29 & 0.09 & 0.26 & 0.28 & 0.08 & Care \\
    Loaded Language & 0.32 & 0.08 & 0.26 & 0.28 & 0.06 & Care \\
    Repetition & 0.30 & 0.09 & 0.26 & 0.30 & 0.05 & Duo \\
    Exaggeration & 0.31 & 0.10 & 0.27 & 0.28 & 0.04 & Care \\
    Obfuscation & 0.28 & 0.10 & 0.27 & 0.27 & \textbf{0.09} & Care \\
    Name Calling & 0.36 & 0.15 & 0.24 & 0.21 & 0.03 & Care \\
    Doubt & 0.35 & 0.13 & 0.22 & 0.26 & 0.04 & Care \\
    Guilt by Association & 0.37 & 0.10 & 0.21 & 0.28 & 0.04 & Care \\
    Appeal to Hypocrisy & 0.34 & 0.11 & 0.26 & 0.26 & 0.04 & Care \\
    \makecell[l]{Questioning \\ the Reputation} & 0.33 & 0.19 & 0.26 & 0.19 & 0.03 & Care \\
    \cmidrule(r){1-7}
    \textbf{Average}& 0.32 & 0.12 & 0.26 & 0.26 & 0.04 & Care \\
    \bottomrule
  \end{tabular}}
\end{table}

The moral resonance of persuasive attacks generated by GPT-4 reveals distinct patterns in how various techniques align with specific moral foundations. Among the techniques analyzed, \textit{Flag Waving} exhibits the strongest connection to \textit{Care} (0.38), underlining its ability to elicit emotional responses by fostering a shared sense of unity and concern. \textit{Causal Oversimplification} stands out with a notable resonance with \textit{Fairness} (0.28), likely because it simplifies cause-and-effect reasoning, provoking concerns related to justice. The highest moral resonance with \textit{Loyalty} is observed in \textit{Appeal to Time} (0.40), emphasizing historical continuity and group traditions, thereby strengthening collective identity and allegiance. \textit{Appeal to Authority}, with the strongest link to \textit{Authority} (0.42), relies on the credibility of figures in positions of power, highlighting the persuasive force of hierarchical trust. While \textit{Purity} resonates less compared to other moral foundations, \textit{Obfuscation} registers the highest resonance with \textit{Purity} (0.09), suggesting that complex or obscure language can generate a sense of exclusivity or sanctity surrounding information.

On average, \textit{Care} is the most prominent moral foundation (0.32), followed by \textit{Authority} (0.26) and \textit{Loyalty} (0.26), with \textit{Fairness} (0.12) and \textit{Purity} (0.04) showing weaker resonance. This implies that GPT-4’s persuasive tactics are primarily structured to trigger emotional responses focused on protection, trust in authority, and group cohesion. However, several attacks deviate from this pattern, prioritizing different moral foundations. For instance, \textit{Appeal to Authority} has a stronger alignment with \textit{Authority} than with \textit{Care}, reinforcing hierarchical deference and reinforcing institutional trust and stability. \textit{Appeal to Time}, conversely, resonates more with \textit{Loyalty} than \textit{Care}, focusing on tradition, continuity, and societal preservation, while emphasizing loyalty to the past and cultural heritage. In addition, \textit{False Dilemma} and \textit{Whataboutism} show significant resonance with \textit{Loyalty}, highlighting group identity appeals and affiliations rather than emotional appeal. Furthermore, \textit{Slogans} and \textit{Repetition} distribute their resonance across multiple moral dimensions rather than focusing on a single foundation, offering broader persuasive influence.

\subsection{Gemma 2 Findings}
\begin{table}
\footnotesize
  \caption{\textbf{Moral resonance of Gemma 2 generated attacks.} These attacks predominantly align with \textit{Care}, highlighting emotional sensitivity and concern for others. \textit{Authority} and \textit{Loyalty} also play significant roles, reinforcing respect for hierarchy and commitment to the group. In contrast, \textit{Fairness} and \textit{Purity} have much weaker resonance, suggesting that appeals to justice and moral purity are less central to its persuasive strategies.}
  \label{gemma2-table}
  \centering
  \resizebox{0.87\textwidth}{!}{
  \begin{tabular}{lcccccc}
    \toprule
    \textbf{Name} & \textbf{Care} & \textbf{Fairness} & \textbf{Loyalty} & \textbf{Authority} & \textbf{Purity} & \textbf{Resonance} \\
    \midrule
    Appeal to Authority & 0.38 & 0.10 & 0.24 & 0.26 & 0.02 & Care \\
    Appeal to Popularity & 0.43 & 0.04 & 0.20 & 0.30 & 0.03 & Care \\
    Appeal to Values & 0.47 & 0.09 & 0.20 & 0.20 & 0.04 & Care \\
    Appeal to Fear & 0.40 & 0.06 & 0.27 & 0.25 & 0.02 & Care \\
    Flag Waving & 0.54 & 0.09 & 0.22 & 0.13 & 0.01 & Care \\
    \makecell[l]{Causal \\ Oversimplification} & 0.43 & 0.10 & 0.24 & 0.20 & 0.03 & Care \\
    False Dilemma & 0.42 & 0.10 & 0.20 & 0.26 & 0.02 & Care \\
    \makecell[l]{Consequential \\ Oversimplification} & 0.34 & 0.06 & 0.23 & 0.33 & 0.03 & Care \\
    Straw Man & 0.34 & \textbf{0.14} & 0.11 & 0.40 & 0.01 & Authority \\
    Red Herring & 0.48 & 0.11 & 0.20 & 0.19 & 0.02 & Care \\
    Whataboutism & 0.35 & 0.13 & 0.24 & 0.24 & 0.04 & Care \\
    Slogans & 0.39 & 0.06 & 0.20 & 0.32 & 0.02 & Care \\
    Appeal to Time & 0.44 & 0.02 & \textbf{0.35} & 0.17 & 0.01 & Care \\
    Conversation Killer & 0.31 & 0.09 & 0.21 & 0.36 & 0.03 & Authority \\
    Loaded Language & 0.33 & 0.07 & 0.19 & 0.37 & 0.05 & Authority \\
    Repetition & 0.25 & 0.07 & 0.14 & \textbf{0.51} & 0.03 & Authority \\
    Exaggeration & \textbf{0.62} & 0.05 & 0.12 & 0.19 & 0.02 & Care \\
    Obfuscation & 0.52 & 0.11 & 0.17 & 0.17 & 0.03 & Care \\
    Name Calling & 0.44 & 0.08 & 0.31 & 0.14 & 0.02 & Care \\
    Doubt & 0.50 & 0.05 & 0.15 & 0.28 & 0.01 & Care \\
    Guilt by Association & 0.43 & 0.03 & 0.07 & 0.45 & 0.02 & Authority \\
    Appeal to Hypocrisy & 0.31 & 0.10 & 0.19 & 0.38 & 0.03 & Authority \\
    \makecell[l]{Questioning \\ the Reputation} & 0.46 & 0.11 & 0.23 & 0.15 & \textbf{0.05} & Care \\
    \cmidrule(r){1-7}
    \textbf{Average} & 0.42 & 0.08 & 0.20 & 0.27 & 0.03 & Care \\
    \bottomrule
  \end{tabular}}
\end{table}

The moral resonance of persuasive attacks generated by Gemma-2 reveals distinct patterns in how different techniques align with specific moral foundations. Among the persuasive techniques analyzed, \textit{Exaggeration} emerges as the most resonant with \textit{Care} (0.62), suggesting its capacity to evoke a strong sense of urgency or concern, particularly in narratives about harm, protection, and emotional appeals. \textit{Straw Man}, with the strongest connection to \textit{Fairness} (0.14), distorts opposing arguments in a way that can spark debates about justice, fairness, and equitable discourse, fostering critical engagement. \textit{Appeal to Time} aligns most closely with \textit{Loyalty} (0.35), reinforcing the importance of tradition, social cohesion, and collective identity within groups. \textit{Repetition}, on the other hand, resonates most with \textit{Authority} (0.51), signifying that repeated assertions bolster trust in authority figures and institutional credibility. Although \textit{Purity} remains the least resonant moral foundation, \textit{Questioning the Reputation} shows the highest connection (0.05), indicating that challenges to credibility may subtly engage concerns about moral purity, integrity, and moral standing.

On average, \textit{Care} is the dominant moral foundation (0.42), followed by \textit{Authority} (0.27), \textit{Loyalty} (0.20), \textit{Fairness} (0.08), and \textit{Purity} (0.03). Most persuasive techniques appeal to emotional responses related to concern and protection, while leveraging hierarchical trust and group allegiance. Certain attacks, such as \textit{Straw Man}, \textit{Conversation Killer}, \textit{Loaded Language}, \textit{Repetition}, \textit{Guilt by Association}, and \textit{Appeal to Hypocrisy}, align more strongly with \textit{Authority}, underscoring their reliance on power. There are differences in moral resonance, despite all aligning with \textit{Authority}. \textit{Guilt by Association} has the highest \textit{Care} score (0.43), showing a stronger appeal to emotional sensitivity, while \textit{Repetition} has the lowest \textit{Care} score (0.25), relying on reinforcing authority. In terms of \textit{Fairness}, \textit{Straw Man} has the highest score (0.14), making a stronger appeal to fairness, while \textit{Guilt by Association} has the lowest (0.03), indicating minimal engagement with justice. For \textit{Loyalty}, \textit{Conversation Killer} leads with a score (0.21), showing moderate appeal to group allegiance, while \textit{Guilt by Association} again has the lowest (0.07), indicating minimal connection to loyalty.

\subsection{Llama 3.1 Findings}
\begin{table}
\footnotesize
  \caption{\textbf{Moral resonance of Llama 3.1 generated attacks.} Attacks  draw on \textit{Loyalty} and \textit{Care}, emphasizing group allegiance, solidarity, and emotional appeal with compassion. \textit{Authority} also plays a notable role, invoking respect for leadership. In contrast, \textit{Fairness} and \textit{Purity} hold  less weight, suggesting that arguments based on justice and moral cleanliness are not as prominent.}
  \label{llama3.1-table}
  \centering
  \resizebox{0.87\textwidth}{!}{
  \begin{tabular}{lcccccc}
    \toprule
    \textbf{Name} & \textbf{Care} & \textbf{Fairness} & \textbf{Loyalty} & \textbf{Authority} & \textbf{Purity} & \textbf{Resonance} \\
    \midrule
    Appeal to Authority & 0.28 & 0.10 & 0.34 & 0.25 & 0.03 & Loyalty \\
    Appeal to Popularity & 0.31 & 0.05 & 0.35 & 0.25 & 0.04 & Loyalty \\
    Appeal to Values & 0.28 & 0.12 & 0.28 & 0.24 & \textbf{0.07} & Duo \\
    Appeal to Fear & 0.31 & 0.08 & 0.34 & 0.26 & 0.02 & Loyalty \\
    Flag Waving & \textbf{0.39} & 0.06 & 0.34 & 0.19 & 0.02 & Care \\
    \makecell[l]{Causal \\ Oversimplification} & 0.31 & \textbf{0.18} & 0.27 & 0.21 & 0.03 & Care \\
    False Dilemma & 0.34 & 0.08 & 0.33 & 0.23 & 0.02 & Care \\
    \makecell[l]{Consequential \\ Oversimplification} & 0.30 & 0.05 & 0.34 & 0.27 & 0.03 & Loyalty \\
    Straw Man & 0.35 & 0.10 & 0.27 & 0.25 & 0.02 & Care \\
    Red Herring & 0.34 & 0.10 & 0.31 & 0.21 & 0.04 & Care \\
    Whataboutism & 0.29 & 0.06 & 0.35 & 0.26 & 0.03 & Loyalty \\
    Slogans & 0.32 & 0.07 & 0.30 & \textbf{0.30} & 0.02 & Care \\
    Appeal to Time & 0.25 & 0.05 & \textbf{0.53} & 0.16 & 0.01 & Loyalty \\
    Conversation Killer & 0.21 & 0.11 & 0.33 & 0.28 & 0.06 & Loyalty \\
    Loaded Language & 0.28 & 0.07 & 0.32 & 0.29 & 0.04 & Loyalty \\
    Repetition & 0.26 & 0.13 & 0.29 & 0.29 & 0.03 & Duo \\
    Exaggeration & 0.36 & 0.06 & 0.32 & 0.22 & 0.03 & Care \\
    Obfuscation & 0.33 & 0.10 & 0.29 & 0.23 & 0.05 & Care \\
    Name Calling & 0.31 & 0.06 & 0.39 & 0.20 & 0.04 & Loyalty \\
    Doubt & 0.34 & 0.09 & 0.27 & 0.27 & 0.03 & Care \\
    Guilt by Association & 0.34 & 0.09 & 0.25 & 0.29 & 0.03 & Care \\
    Appeal to Hypocrisy & 0.26 & 0.11 & 0.28 & 0.30 & 0.04 & Authority \\
    \makecell[l]{Questioning \\ the Reputation} & 0.33 & 0.15 & 0.29 & 0.18 & 0.05 & Care \\
    \cmidrule(r){1-7}
    \textbf{Average} & 0.31 & 0.09 & 0.32 & 0.24 & 0.03 & Loyalty \\
    \bottomrule
  \end{tabular}}
\end{table}

The moral resonance of persuasive attacks generated by Llama3.1 reveals distinct patterns in how different techniques align with specific moral foundations. Among the persuasive techniques analyzed, \textit{Flag Waving} resonates most strongly with \textit{Care} (0.39), underlining its emotional appeal tied to concern for well-being, as it often unites individuals under a shared cause and promotes collective empathy and social cohesion. \textit{Causal Oversimplification} shows the greatest resonance with \textit{Fairness} (0.18), as its oversimplified reasoning often raises issues of justice and fairness, provoking moral reflection and ethical consideration. The most significant connection to \textit{Loyalty} is found in \textit{Appeal to Time} (0.53), emphasizing historical continuity and traditional values, central to fostering group loyalty, solidarity, and unity. In terms of \textit{Authority}, \textit{Slogans} exhibit the highest resonance (0.30), as their repetitive nature lends credibility to institutional or authoritative figures. 

On average, \textit{Loyalty} emerges as the dominant moral foundation (0.32), followed closely by \textit{Care} (0.31), and \textit{Authority} (0.24), with \textit{Fairness} (0.09) and \textit{Purity} (0.03) trailing behind. This indicates that many of these attacks are primarily geared toward reinforcing group identity, tradition, and emotional concerns for protection. \textit{Care} also plays an important role, reflecting emotional appeals focused on concern, well-being, and empathy. While most persuasive attacks mainly align with both \textit{Loyalty} and \textit{Care}, seven attacks stand out by strongly incorporating \textit{Authority}: \textit{Slogans}, \textit{Conversation Killer}, \textit{Loaded Language}, \textit{Repetition}, \textit{Doubt}, \textit{Guilt by Association}, and \textit{Appeal to Hypocrisy}. These attacks emphasize leadership, structure, and deference to norms alongside their persuasive elements. For instance, \textit{Slogans} and \textit{Appeal to Hypocrisy} (0.30) rely heavily on institutional credibility, while \textit{Conversation Killer} (0.28), \textit{Loaded Language} (0.29), and \textit{Doubt} (0.27) use authority-driven rhetoric to shape and direct discourse. \textit{Repetition} (0.29) and \textit{Guilt by Association} (0.29) reinforce ideological consistency and collective identity among group members.

\subsection{Comparative Findings}
The moral resonance of persuasive attacks generated by GPT-4, Gemma 2, and Llama 3.1 reveals shared trends and distinct patterns in how different techniques align with moral foundations. All three models focus significantly on \textit{Care}, though to varying extents. Gemma 2 stands out with a notably higher resonance for \textit{Care} (0.42), particularly through techniques like \textit{Exaggeration} (0.62), which evokes emotional urgency and concern. GPT-4 and Llama 3.1 also show considerable resonance with \textit{Care}, with scores of 0.32 and 0.31, respectively. While \textit{Care} is prominent in all three models, GPT-4 and Llama 3.1 both give greater weight to \textit{Loyalty}, with Llama 3.1 placing the strongest emphasis on \textit{Loyalty} (0.32) through techniques like \textit{Appeal to Time} (0.53), reinforcing tradition and group allegiance. GPT-4’s \textit{Appeal to Time} (0.40) also underscores group cohesion. On the other hand, Gemma 2, while focusing on \textit{Care}, shows a significant connection to \textit{Authority} (0.27), especially through \textit{Repetition} (0.51), highlighting hierarchical trust. This contrasts with Llama 3.1, where \textit{Authority} is less central, resonating more with \textit{Loyalty} and \textit{Care}.

The key difference between the models lies in their approach to aligning persuasive attacks with the moral foundations of \textit{Care}, \textit{Loyalty}, and \textit{Authority}. For \textit{Care}, GPT-4’s \textit{Flag Waving} resonates moderately (0.38), focusing on unity, but it is less pronounced than Gemma 2's \textit{Exaggeration}, which shows strong resonance (0.62), evoking urgency and emotional appeals related to harm and protection. Llama 3.1 also aligns \textit{Flag Waving} with \textit{Care} (0.39), emphasizing collective empathy and cohesion, though with a slightly weaker resonance compared to Gemma 2. In \textit{Loyalty}, all three models prioritize \textit{Appeal to Time}, but Llama 3.1 has the highest resonance (0.53), highlighting historical continuity and group traditions. GPT-4 follows (0.40), and Gemma 2 aligns it (0.35), reinforcing group solidarity. In \textit{Authority}, Gemma 2’s \textit{Repetition} leads with the highest resonance (0.51), reflecting the power of repeated assertions in building trust. GPT-4’s \textit{Appeal to Authority} follows (0.42), leveraging the credibility of authoritative figures, while Llama 3.1's \textit{Slogans}, with a resonance (0.30), use authoritative voices moderately to effectively deliver the message.

\section{Discussion}
The findings from GPT-4, Gemma 2, and Llama 3.1 generated persuasive attacks extend previous research on the persuasive capabilities of LLMs \cite{garry2024large}. Prior work has shown that LLMs can generate narratives that frame discussions in particular ways, influencing how information is perceived and internalized \cite{williams2024large}. The observed alignment of persuasive techniques with moral foundations suggests that these models do more than simply generate content—they actively structure communication to evoke emotional and cognitive responses. This resonates with earlier research on how persuasive messaging can be strategically designed to reinforce specific viewpoints while minimizing alternative perspectives \cite{grovs2024information}. The ability of LLMs to generate persuasive content at scale underscores their influence on collective intelligence, as their generated narratives may dominate public discussions and subtly shift the moral framing of key issues \cite{burton2024large}. These findings further underscore the importance of examining how LLMs shape discourse through persuasive attacks.

The moral resonance of these models highlights key patterns in how persuasive techniques align with fundamental human values \cite{feinberg2019moral}. Prior studies have shown that persuasive appeals are most effective when they connect with deeply held moral beliefs, and the findings from these models reflect this dynamic across different persuasive strategies \cite{kodapanakkal2022moral}. The emphasis on \textit{Care}, \textit{Loyalty}, and \textit{Authority} across models suggests that LLM-generated persuasion often mirrors real-world rhetorical strategies that leverage emotional connection, group cohesion, and trust in authoritative figures \cite{luttrell2019challenging}. These moral foundations have been extensively studied in persuasion theory, particularly in how they shape resistance to opposing views \cite{harre1985persuasion}. The findings also align with research on online persuasion, where techniques such as repetition and emotional framing are known to amplify certain narratives while marginalizing dissenting perspectives \cite{chen2021persuasion}. This work advances discussions on how LLMs are transforming the information landscape through their persuasive capabilities.

The comparative analysis of GPT-4, Gemma 2, and Llama 3.1 reveals both shared tendencies and distinct differences in how these models structure persuasive messaging, extending persuasion research \cite{dimitrov2021semeval}. All three models align with moral foundations related to emotional and group-based appeals, reinforcing findings that persuasion relies on eliciting concern, fostering social cohesion, and establishing credibility through authority \cite{feinberg2019moral}. However, their approaches vary in moral resonance, with GPT-4 emphasizing emotional appeals for protection and well-being, balanced by trust and cohesion, Gemma 2 prioritizing urgency with trust and group allegiance, and Llama 3.1 focusing on emotional appeals and group loyalty rooted in tradition. These differences suggest that while LLMs use the same persuasive techniques, their moral framing can differ, highlighting how model-specific tendencies influence the ethical dimensions of persuasive outputs. This additional consideration to persuasive techniques is not fully addressed in previous work \cite{piskorski2023semeval}. By gaining awareness of how LLMs strategically use persuasive techniques, individuals and organizations can become more attuned to the subtle ways these models shape narratives and influence public discourse. 

\paragraph{Limitations}
There are potential biases inherent in the LLMs used for generating attacks, such as GPT-4, Gemma 2, and Llama 3.1. These models are trained on diverse datasets that may reflect societal, political, or cultural biases, influencing the persuasive attacks they generate. While these biases may shape the moral resonance and techniques of the attacks, the dataset provides a foundation for exploring these dynamics further. The concept of moral resonance, which aligns persuasive attacks with moral appeals like \textit{Care}, \textit{Authority}, and \textit{Loyalty}, may not always match real-world outcomes, as the impact of these appeals can vary depending on interpretation, context, and circumstances. Additionally, variability in context, such as the tone of the targeted news releases, may influence the form of the persuasive attacks. These nuances highlight the importance of a large LLM-generated persuasion attack dataset, offering an opportunity for future researchers to explore how contextual factors and model biases interact in persuasive strategies. By providing a rich resource of LLM-generated persuasive content, this paper sets the stage for continued investigation into these research questions.

\paragraph{Ethical Considerations}
This initiative aims to support efforts to navigate the evolving and complex information landscape shaped by the emergence of LLMs. The primary objective is to create a large-scale dataset consisting of comprehensive LLM-generated persuasive attacks toward government agency news, sourced from a diverse range of articles across ten different government agencies. We recognized that the most ethical risk comes from revealing the attack generation prompt, therefore, we only provide a high-level overview of the prompt structure to prevent this while still offering sufficient technical discussion. This dataset will serve as a valuable resource, enabling agencies, individuals, and organizations to see how LLMs can efficiently drive competing narratives on critical topics that affect everyday lives. It will offer valuable research opportunities to better understand the complex mechanisms behind LLM-driven persuasive attacks. Our analysis of the moral resonance of LLM-generated persuasive attacks highlights the potential of this dataset to reveal the underlying strategies behind these attacks, marking a critical step toward understanding their persuasiveness.

\section{Summary and Operational Implications}
The era of generative AI has fundamentally reshaped the information ecosystem, affecting not just organizations but also individuals and communities. As these models generate persuasive content at scale, they create an uphill battle for effective and resilient communication for their human counterparts, given the vast difference in capabilities between human and LLM-driven communicators. This dataset provides significant research opportunities to study LLM-generated persuasive attacks that address key societal issues. For our analysis, we focused on analyzing the moral resonance of these attacks, which contributes to the overall persuasive signature of LLM models and persuasive techniques. These can be expanded to socio-emotional-cognitive signatures e.g., subjectivity, toxicity, emotions, attitudes, intent etc.~\cite{Volkova2021}.

This large LLM-generated persuasion attack dataset not only advances academic understanding but also has significant operational implications for agencies' communication strategies. By providing unprecedented insight into how LLMs construct persuasive attacks across various techniques and models, it enables organizations to develop targeted countermeasures and more resilient messaging frameworks. This operational advantage allows agencies to proactively address potential vulnerabilities in their communications, anticipate attack vectors based on moral resonance patterns, and design messaging that maintains integrity even when subjected to AI-generated criticism. As information operations increasingly leverage LLMs, this dataset serves as a critical resource for building organizational "reputation armor" that can withstand sophisticated persuasion attempts, ultimately strengthening public trust in official communications across contested information environments.

\begin{ack}
Distribution Statement A – Approved for Public Release, Distribution Unlimited.
\end{ack}



\newpage
\bibliographystyle{acm}
\bibliography{Reference}

\end{document}